\documentclass[reprint,onecolumn,amsmath,amssymb,superscriptaddress,jcp]{revtex4}
\usepackage{mathtools}
\usepackage{esint}
\usepackage{graphicx}
\usepackage{graphics}
\usepackage{multirow}
\usepackage{bm}
\usepackage{listings}
\usepackage{color}
\usepackage[caption=false]{subfig}
\usepackage[colorlinks]{hyperref}
\usepackage{algorithm}
\usepackage{algpseudocode}
\algblock{Input}{EndInput}
\algnotext{EndInput}
\algblock{Output}{EndOutput}
\algnotext{EndOutput}
\algblock{Filtering}{EndFiltering}
\algnotext{EndFiltering}
\algblock{Projection}{EndProjection}
\algnotext{EndProjection}
\algblock{Diagonalization}{EndDiagonalization}
\algnotext{EndDiagonalization}
\algblock{Rotation}{EndRotation}
\algnotext{EndRotation}

\makeatletter
\newcommand*{\rom}[1]{\expandafter\@slowromancap\romannumeral #1@}
\makeatother

\newlength{\bracewidth}
\newcommand{\myunderbrace}[2]{\settowidth{\bracewidth}{$#1$}#1\hspace*{-1\bracewidth}\smash{\underbrace{\makebox{\phantom{$#1$}}}_{#2}}}

\begin{document}
\title{GPU acceleration of local and semilocal density functional calculations in the SPARC electronic structure code}

\author{Abhiraj Sharma}
\affiliation{Physics Division, Lawrence Livermore National Laboratory, Livermore, CA, 94550, USA}
\author{Alfredo Metere}
\affiliation{Physics Division, Lawrence Livermore National Laboratory, Livermore, CA, 94550, USA}
\author{Phanish Suryanarayana}
\affiliation{College of Engineering, Georgia Institute of Technology, Atlanta, Georgia 30332, USA}
\author{Lucas Erlandson}
\affiliation{College of Computing, Georgia Institute of Technology, Atlanta, Georgia 30332, USA}
\author{Edmond Chow}
\affiliation{College of Computing, Georgia Institute of Technology, Atlanta, Georgia 30332, USA}
\author{John E. Pask}
\email[Email: ]{pask1@llnl.gov}
\affiliation{Physics Division, Lawrence Livermore National Laboratory, Livermore, CA, 94550, USA}
%%%%%%%%%%%%%%%%%%%%%%%%%%%%%%%%%%%%%%%%%%%%%%%%%%%%%%%%%%%%%%%%%%%%%%%%%%%%%%%%%%%%%%%%%%%%%%%%%%%%%%%%%%%%%%%%%%%%%%%%%%%%%%%%%%%%%%%%%%%%%%%%%%%%%%%%%%%%%%%%%%%%%%%%%%%%%%%%%%%%%%%%%%%%%%%%%%%%
\begin{abstract}
We present a GPU-accelerated version of the real-space SPARC electronic structure code for performing Kohn-Sham density functional theory calculations within the local density and generalized gradient approximations. In particular, we develop a modular math kernel based implementation for \texttt{NVIDIA} architectures wherein the computationally expensive operations are carried out on the GPUs, with the remainder of the workload  retained on the CPUs. Using representative bulk and slab examples, we show that GPUs enable speedups of up to 6x relative to CPU-only execution, bringing time to solution down to less than 30 seconds for a metallic system with over 14,000 electrons, and enabling significant reductions in computational resources required for a given wall time.
\end{abstract}
\maketitle
\allowdisplaybreaks
%%%%%%%%%%%%%%%%%%%%%%%%%%%%%%%%%%%%%%%%%%%%%%%%%%%%%%%%%%%%%%%%%%%%%%%%%%%%%%%%%%%%%%%%%%%%%%%%%%%%

\section{Introduction}

Over the past few decades, Kohn-Sham density functional theory (DFT) \cite{Hohenberg, Kohn1965} has established itself as one of the cornerstones of materials and chemical sciences research. In particular, due to its high accuracy-to-cost ratio relative to other ab initio methods, it has seen widespread use for understanding as well as  predicting material properties and chemical phenomena from the first principles of quantum mechanics \cite{burke2012dft, becke2014perspective}. In spite of significant advances, in numerical/computational algorithms as well as high-performance computing architectures, bringing down the time to  solution of the Kohn-Sham problem remains a challenging task. In particular, the computational cost and memory requirements scale cubically and quadratically with system size, respectively, restricting the range and types of systems that can be investigated, particularly in ab-initio molecular dynamics (AIMD) simulations, wherein reaching time scales of interest might necessitate the solution of the Kohn-Sham equations tens or hundreds of thousands of times \cite{burke2012dft}.

The planewave pseudopotential method \cite{Martin2004}, which employs the complete, orthogonal, Laplacian-diagonalizing, periodic, and atom position independent   Fourier basis for discretization, is among the most widely used techniques for the solution of the Kohn-Sham equations \cite{VASP, CASTEP, ABINIT, Espresso, CPMD, DFT++, gygi2008architecture, valiev2010nwchem}. In particular, the planewave method is accurate, relies on a single parameter for convergence with basis, and is highly efficient on small to moderate computational resources through the use of efficient preconditioning schemes and well optimized Fast Fourier Transforms (FFTs). However, the planewave method is restricted to periodic boundary conditions, wherein artificial periodicity has to be introduced through large vacuum regions for systems that are finite in one or more directions.  Moreover, the global nature of the Fourier basis makes the development of linear-scaling methods \cite{Goedecker, Bowler2012, aarons2016perspective} difficult, and limits the parallel scalability of the planewave method on large-scale computational resources, which severely restricts the system sizes and time scales accessible to a rigorous first-principles Kohn-Sham DFT investigation.

Motivated by the limitations of the planewave method, a number of alternate solution strategies based on systematically improvable, localized representations have been developed \cite{becke1989basis, chelikowsky1994finite, genovese2008daubechies, seitsonen1995real, white1989finite, iwata2010massively, tsuchida1995electronic, xu2018discrete, Phanish2011, Phanish2010, ONETEP, CONQUEST, das2022dft, OCTOPUS, briggs1996real, fattebert1999finite, shimojo2001linear, Ghosh2017extended, arias1999wav, pask2005femeth, lin2012adaptive}, among which the real-space finite-difference method \cite{beck2000rsmeth,saad2010esmeth} is perhaps the most mature and widely used to date. In this method, computational locality is maximized by discretizing all spatial quantities on a uniform, atom position independent real-space grid, wherein convergence is controlled by a single parameter, i.e., the grid spacing. The method naturally accommodates both periodic and Dirichlet boundary conditions, allowing for the accurate and efficient treatment of systems with different dimensionalities, i.e., finite, semi-infinite, and bulk, and even those with non-traditional symmetries \cite{sharma2021real, ghosh2019symmetry}. Moreover, the localized real-space representation allows for the development of linear scaling methods, and being free from communication-intensive transforms such as FFTs, the method allows for large-scale parallel computational resources to be efficiently leveraged  \cite{shimojo2001linear, iwata2010massively, hasegawa2011first, osei2014accurate, suryanarayana2018sqdft, gavini2022roadmap}.

SPARC \cite{xu2021sparc, Ghosh2017extended, Ghosh2017cluster} is a recently developed open source electronic structure code that incorporates a number of the developments in real-space DFT made over the past decade, allowing for efficient utilization of modest as well as large-scale computational resources.  Its accuracy and performance have been extensively verified and benchmarked against established planewave codes, during which it has been found to be an order of magnitude faster for local, semilocal, and hybrid exchange-correlation functionals, with increasing advantages as the number of processors is increased \cite{xu2021sparc, gavini2022roadmap}. However, it has heretofore been unable to exploit the acceleration provided by Graphics Processing Units (GPUs), which have been shown to provide substantial speedups in the context of electronic structure calculations \cite{walker2016electronic, gonze2016recent, genovese2009density, genovese2016wavelet, manninen2013applied, maintz2011speeding, hacene2012accelerating, jia2017gpu, andrade2012time, wilkinson2013porting, jia2013fast,romero2018performance, huhn2020gpu, das2022dft}, providing the motivation for the current work. In particular, we develop a modular math kernel based GPU-accelerated version of SPARC for local and semilocal Kohn-Sham DFT calculations, wherein the computationally expensive operations are carried out on the GPUs, with the remainder of the workload  retained on the CPUs. Using representative bulk and slab examples, we show that GPU-acceleration provides speedups of up to 6x, bringing time to solution down to less than 30 seconds for a metallic system with over 14,000 electrons, and enabling significant reductions in computational resources required for a given wall time.

The remainder of this paper is organized as follows. In Section~\ref{Sec:GPUaccel}, we describe the GPU acceleration of local/semilocal DFT calculations in SPARC. Next, we verify the performance of the GPU accelerated SPARC code in Section~\ref{Sec:Results}. Finally, we provide concluding remarks in Section~\ref{Sec:Conclusions}.

%%%%%%%%%%%%%%%%%%%%%%%%%%%%%%%%%%%%%%%%%%%%%%%%%%%%%%%%%%%%%%%%%%%%%%%%%%%%%%%%%%%%%%%%%%%%%%%%%%%%

\section{GPU acceleration of local and semilocal DFT calculations in SPARC} \label{Sec:GPUaccel}

The electronic ground state in SPARC \cite{xu2021sparc, Ghosh2017extended, Ghosh2017cluster} is determined using the self-consistent field (SCF) method \cite{Martin2004}, which represents a fixed-point iteration with respect to either the density or potential. In each SCF iteration, a Schr{\"o}dinger-type linear eigenproblem is solved for the eigenvectors/orbitals and the Poisson equation is solved for the electrostatic potential. Given the large prefactor and $\mathcal{O}(N^3)$ scaling with system size, the overall computational cost of Kohn-Sham DFT calculations is primarily determined by the solution of the eigenproblem, especially when the exchange-correlation functional is approximated using either the local density  approximation (LDA) or generalized gradient approximation (GGA) \cite{Martin2004}, which is the focus of the current work.  

SPARC employs the Chebyshev-filtered subspace iteration (CheFSI) \cite{zhou2006self, zhou2006parallel} to perform partial diagonalization of the Hamiltonian during each SCF iteration, as summarized in Algorithm~\ref{alg:CheFSI}. The CheFSI algorithm consists of two main steps, namely Chebyshev filtering and Rayleigh-Ritz. In Chebyshev filtering, the rapid growth of Chebyshev polynomials outside the interval [-1,1] is used to filter out the unwanted part of the Hamiltonian's spectrum, i.e., the unoccupied subspace. In Rayleigh-Ritz --- which consists of projection of the Hamiltonian onto the filtered subspace, diagonalization of the resulting subspace Hamiltonian, and rotation of the filtered basis  --- approximations to the eigenvectors and eigenvalues of the Hamiltonian are then calculated. Indeed, as the SCF iteration proceeds towards self-consistency, these eigenvectors converge to the Kohn-Sham orbitals. 

%%%%%%%%%%%%%%%%%%%%%%%%%
 \begin{algorithm}[h!]
	\caption{CheFSI-based partial diagonalization} \label{alg:CheFSI}
	\begin{algorithmic}[]
	\item $H:$ Hamiltonian for a given spin, Bloch wavevector, and  SCF iteration, a matrix of size $N_d \times N_d$ applied as an operator
	\item $X:$ guess for the eigenvectors/orbitals, a matrix of size $N_d \times N_s$
	\item $N_d:$ number of finite-difference nodes, $N_s$: number of orbitals \vspace{2mm}
       
        \Filtering
        \begin{itemize}
         \item Enhance desired part of the spectrum through Chebyshev polynomial filtering:
         \begin{align}
      \hspace{-50mm} \widetilde{X} = p_m\bigg(\frac{H-c I}{e}\bigg) X  \,, \nonumber 
       \end{align}
       \hspace{4mm} $p_m$: Chebyshev polynomial of degree $m$, $c = (\lambda_{N_d} + \lambda_c)/2$, $e = (\lambda_{N_d} - \lambda_c)/2$ \,,
       \newline \hspace*{4mm} $\lambda_{N_d}$: largest eigenvalue of $H$,  $\lambda_c$: filter cutoff. 
       \item Key computational kernel and its scaling: 
       \begin{align}
       \hspace{-50mm} HX  \,, \quad \mathcal{O}(N_d N_s) \,. \nonumber
       \end{align}
        \end{itemize}
      \EndFiltering

     \Projection
     \begin{itemize}
     \item Project Hamiltonian onto the Chebyshev filtered basis:  
     \begin{align}
    \hspace{-50mm}  \widetilde{H} = \widetilde{X}^T H \widetilde{X} \,, \quad \widetilde{M} = \widetilde{X}^T \widetilde{X} \,. \nonumber
     \end{align}
     \item Key computational kernels and their scaling: 
       \begin{align}
      \hspace{-35mm}  H \widetilde{X} \,, \quad & \mathcal{O}(N_d N_s) \,, \nonumber  \\
     \hspace{-35mm}    \widetilde{X}^T \widetilde{Y}  \,, \quad & \mathcal{O}(N_d N_s^2) \,,  \nonumber \\
     \hspace{-35mm}    \widetilde{X}^T \widetilde{X}  \,, \quad & \mathcal{O}(N_d N_s^2) \,, \nonumber
       \end{align}
      \hspace{4mm}  $\widetilde{Y}$: matrix of size $N_d\times N_s$. 
     \end{itemize}
        \EndProjection 
       
 \Diagonalization
 \begin{itemize}
 \item Solve subspace eigenproblem: 
 \begin{align}
 \hspace{-50mm}  \widetilde{H} \widetilde{Z} =  \widetilde{M} \widetilde{Z}  \widetilde{D}  \,. \nonumber 
 \end{align}
  \hspace{4mm} $\widetilde{Z}$: matrix of size $N_s \times N_s$ 
  \newline \hspace*{4mm}   $\widetilde{D}$: Diagonal matrix of size $N_s \times N_s$
\item Key computational kernel and its scaling: 
\begin{align}
 \hspace{-50mm} \text{Eigendecomposition}\,, \quad \mathcal{O}(N_s^3) \,. \nonumber 
\end{align}
 \end{itemize}
        \EndDiagonalization
        
        \Rotation
        \begin{itemize}
        \item Subspace rotation step to obtain approximate eigenvectors of $H$, used as the guess for the next SCF step: 
        \begin{align}
        \hspace{-50mm}  X = \widetilde{X} \widetilde{Z} \,. \nonumber
        \end{align}
        \item Key computational kernel and its scaling: 
        \begin{align}
         \hspace{-50mm} \widetilde{X} \widetilde{Z} \,, \quad \mathcal{O}(N_d N_s^2) \,. \nonumber 
        \end{align}
        \end{itemize}
        \EndRotation
	\end{algorithmic} 
\end{algorithm}

%%%%%%%%%%%%%%%%%%%%%%%%%%%%%%

In the CPU implementation of SPARC, parallelization is achieved using the Message Passing Interface (MPI) standard. In particular,  an eigensolver topology is implemented for CheFSI in which the \texttt{MPI\_COMM\_WORLD} communicator is split into two spin groups, then each spin group is split into multiple Bloch wavevector groups, then each wavevector group is split into multiple orbital groups, and finally, each orbital group is embedded with a Cartesian topology \cite{xu2021sparc}. In the current GPU-accelerated implementation, we neglect spin and employ only wavevector and orbital parallelization, i.e.,  no domain decomposition, which translates to each orbital group no longer being embedded with a Cartesian topology. Note that it is relatively straightforward to include spin polarization, given that the eigenproblems for different spins are essentially independent. Also note that in the default SPARC operation, the parallelization over all the orbitals occurs first, and then only domain decomposition is activated, i.e., domain decomposition is important in the strong scaling limit, but not in regular operation where moderate number of processors are used, motivating the current choice. 

In this work, we propose a strategy that ensures maximum transferability across diverse and ever-evolving GPU architectures and their corresponding programming interfaces, code separation of CPU and GPU CheFSI modules that allows their independent development and optimizations, minimum data transfer between host and device, and minimum peak memory requirement on a GPU. In what follows, we describe how the key computational kernels in each of the aforementioned CheFSI steps are accelerated on \texttt{NVIDIA}  GPUs using the \texttt{cuBLAS}  and \texttt{cuSOLVER} libraries, via the \texttt{CUDA} parallel programming platform. The vectors/matrices are transferred from the CPU to GPU and GPU to CPU using the \texttt{cublasSetVector} and \texttt{cublasGetVector} routines, respectively. Note that for isolated systems/$\Gamma-$point calculations, real-valued computations are performed, whereas for other choices of Brillouin zone integration, complex-valued computations are performed, with all operations performed in double-precision arithmetic. 

While the current implementation works for any integer CPU-thread-to-GPU ratio greater than or equal to 1, for simplicity of discussion, we assume a CPU-thread-to-GPU ratio  of 1, which is the default and most efficient setting in our implementation, providing an efficient load distribution with minimum PCI bus transactions between the host CPU and mapped device GPU. In addition, we consider a single wavevector in the Brillouin zone, since the parallelization over the different wavevectors  follows naturally, given that the eigenproblems appearing at different wavevectors are essentially independent in the current context. The corresponding Hamiltonian at a given SCF iteration, which is a sparse matrix of $N_d \times N_d$, will be denoted by $H$, and the guess for its eigenvectors/orbitals, which is a dense  matrix of  size $N_d \times N_s$, will be denoted by $X$, where $N_d$ denotes the number of finite-difference nodes and $N_s$ denotes the number of orbitals. We will consider two partitions for $X$ and related quantities:
\begin{align}
X := 
\begin{bmatrix} 
X_{(1)} & X_{(2)}  & \ldots & X_{(p)}
\end{bmatrix} \,, \quad \quad & \text{column-wise partition}  \,, \\
X := \begin{bmatrix}
X^{(1)} \\
X^{(2)} \\
\vdots \\
X^{(p)}
\end{bmatrix} \,, \quad \quad & \text{row-wise partition} \,,
\end{align} 
where $X_{(k)}$ and $X^{(k)}$ are matrices of size $N_d \times N_{s}/p$ and $N_{d}/p \times N_{s}$, respectively, that are  associated with CPU$_{\rm k}$/GPU$_{\rm k}$ and $p$ is the number of CPUs/GPUs. Indeed, if $N_d$ and $N_s$ are not integer multiples of the number of processors, the number of rows and columns for the $p$-th processor are reduced, respectively, such that the sizes of the matrices are the same on the remaining processors. Henceforth, we will use under and side braces to denote which CPU/GPU the matrix resides in, and therefore where the computations are performed (if any).

It is worth noting that some of the routines developed for the implementation of Chebyshev filtering  (Section~\ref{Subsec:ChebFilt}), i.e., stencil operations and  nonlocal projector multiplications can be used to accelerate the computation of the nonlocal component of the Hellmann-Feynman atomic forces and stresses in SPARC, where the key computational kernels are the application of the gradient operator on the Kohn-Sham orbitals, and then the application of the nonlocal pseudopotential operator on the resultant quantity \cite{Ghosh2017extended, Ghosh2017cluster, sharma2018calculation}. For the stresses, the gradient of the orbitals so computed can be used for the calculation of the   electronic kinetic energy component of the stress. Indeed, these nonlocal components of the forces and stresses are explicitly dependent on the orbitals and therefore significantly more expensive than the local components, motivating acceleration through GPU computations.  

%%%%%%%%%%%%%%%%%%%%%%%%%%%%%%%%%%%%%%%%%%%%%

\subsection{Chebyshev filtering} \label{Subsec:ChebFilt}
The guess for the eigenvectors $X$ is initially distributed on the CPU threads as follows:
\begin{align}
X := 
\begin{bmatrix} 
\myunderbrace{X_{(1)}}{\text{CPU}_{1}} & \myunderbrace{X_{(2)}}{\text{CPU}_2}  & \ldots & \myunderbrace{X_{(p)}}{\text{CPU}_{\rm p}} 
\end{bmatrix} \,. \\ 
\nonumber
\end{align} 
The matrix $X$, effective potential, and  nonlocal projectors are then  transferred from the host CPU to its mapped GPU device. The key computational kernel within the Chebyshev filtering is computed as: 
\begin{align}
H X := H \begin{bmatrix} \myunderbrace{X_{(1)}}{\text{GPU}_{1}} & \myunderbrace{X_{(2)}}{\text{GPU}_2}  & \ldots & \myunderbrace{X_{(p)}}{\text{GPU}_{\rm p}}  \end{bmatrix} =  \begin{bmatrix} \myunderbrace{H X_{(1)}}{\text{GPU}_1} & \myunderbrace{H X_{(2)}}{\text{GPU}_2} & \ldots & \myunderbrace{H X_{(p)}}{\text{GPU}_{\rm p}} \end{bmatrix} \,. \\
\nonumber
\end{align}
where the Hamiltonian $H$, which consists of the Laplacian, effective potential (sum of the electrostatic and exchange-correlation potentials), and outer product of the nonlocal projectors,  is never explicitly created, but rather its application on vectors/matrices is computed in matrix-free fashion as follows:

\begin{itemize}
\item The application of the finite-difference stencil for the Laplacian on each column of $X_{(k)}$ by GPU$_{\rm k}$, $k \in \{1, 2, \ldots, p \}$, proceeds as follows \cite{micikevicius20093d}: (i) group threads into 2D threadblocks of size $(p,q)$ to match data tiling in $(x,y)$ and assign one thread per output element; (ii) allocate a shared memory for $(p+n_0) \times (q+n_0)$ array, $n_0$ being the finite-difference order; (iii) load the column of $X_p$ in the shared memory; (iv) compute 2D stencil in each threadblock by fetching data from the shared memory; and (v) compute 1D stencil in z-direction in each threadblock and add to the 2D stencil result. This algorithm ensures minimum read redundancy by collecting the data corresponding to the extended region in the $(x,y)$ tile in the shared memory of a threadblock. In addition, all GPU threads work in parallel, each performing only $3n_0 + 1$ computations, thus enabling very fast and accurate stencil computations.
\item The effective potential is multiplied pointwise to each column of $X_{(k)}$ by GPU$_{\rm k}$, $k \in \{1, 2, \ldots, p \}$. 
\item The nonlocal projectors for each atom, which are stored as a dense matrix, are applied on the appropriate components of $X_{(k)}$ by GPU$_{\rm k}$, $k \in \{1, 2, \ldots, p \}$ by performing a dense matrix-matrix multiplication using the \texttt{cublasZgemm/cublasDgemm} routine.
\end{itemize}
Once the filtering is complete, the filtered basis $\widetilde{X}$ is transferred from GPUs to CPUs. Note that since $\widetilde{Y} = H \widetilde{X}$ is needed as part of the projection step, it is calculated as described above and also transferred from the GPUs to CPUs. 

%%%%%%%%%%%%%%%%%%%%%%%%%%%%%%

\subsection{Projection}

The matrices $\widetilde{X}$ and $\widetilde{Y}$ are first redistributed from a 1D column block distribution to a 1D row block distribution on the CPUs as follows:
\begin{align}
\widetilde{X}: \begin{bmatrix} \myunderbrace{\widetilde{X}_{(1)}}{\text{CPU}_{1}} & \myunderbrace{\widetilde{X}_{(2)}}{\text{CPU}_2}  & \ldots & \myunderbrace{\widetilde{X}_{(p)}}{\text{CPU}_{\rm p}}  \end{bmatrix}  
\longrightarrow
\begin{bmatrix}
\widetilde{X}^{(1)} \\
\widetilde{X}^{(2)} \\
\vdots \\
\widetilde{X}^{(p)}
\end{bmatrix} \hspace{-2mm}
\begin{array}{l}
\} \text{\scriptsize CPU}_{1} \\
\} \text{\scriptsize CPU}_{2} \\
\vspace{3mm}  \\
\} \text{\scriptsize CPU}_{\rm p} \\
\end{array} \,, 
\\ \nonumber \\
\widetilde{Y}: \begin{bmatrix} \myunderbrace{\widetilde{Y}_{(1)}}{\text{CPU}_{1}} & \myunderbrace{\widetilde{Y}_{(2)}}{\text{CPU}_2}  & \ldots & \myunderbrace{\widetilde{Y}_{(p)}}{\text{CPU}_{\rm p}}  \end{bmatrix}  
\longrightarrow
\begin{bmatrix}
\widetilde{Y}^{(1)} \\
\widetilde{Y}^{(2)} \\
\vdots \\
\widetilde{Y}^{(p)}
\end{bmatrix} \hspace{-2mm}
\begin{array}{l}
\} \text{\scriptsize CPU}_{1} \\
\} \text{\scriptsize CPU}_{2} \\
\vspace{3mm}  \\
\} \text{\scriptsize CPU}_{\rm p} \\
\end{array} \,.
\end{align}
Next, $\widetilde{X}$ and $\widetilde{Y}$ are transferred from the CPUs to the GPUs, after which the subspace Hamiltonian $\widetilde{H}$ and overlap $\widetilde{M}$ matrices are computed as follows: 
\begin{align}
\widetilde{H} = \widetilde{X}^{T} \widetilde{Y} := 
\begin{bmatrix} \myunderbrace{\widetilde{X}^{{(1)}^T}}{\text{GPU}_{1}} & \myunderbrace{\widetilde{X}^{{(2)}^T}}{\text{GPU}_2}  & \ldots & \myunderbrace{\widetilde{X}^{{(p)}^T}}{\text{GPU}_{\rm p}}  \end{bmatrix}  
\begin{bmatrix}
\widetilde{Y}^{(1)} \\
\widetilde{Y}^{(2)} \\
\vdots \\
\widetilde{Y}^{(p)}
\end{bmatrix} \hspace{-2mm}
\begin{array}{l}
\} \text{\scriptsize GPU}_{1} \\
\} \text{\scriptsize GPU}_{2} \\
\vspace{3mm}  \\
\} \text{\scriptsize GPU}_{p} \\
\end{array} 
=
[
\underbrace{\widetilde{X}^{{(1)}^T} \widetilde{Y}^{(1)}}_{\text{GPU}_1} + 
\underbrace{\widetilde{X}^{{(2)}^T} \widetilde{Y}^{(2)}}_{\text{GPU}_2} + 
\ldots +
\underbrace{\widetilde{X}^{{(p)}^T} \widetilde{Y}^{(p)}}_{\text{GPU}_{\rm p}}
]  \nonumber \\
:= [
\underbrace{\underbrace{\widetilde{H}^{(1)}}_{\text{CPU}_1} + 
\underbrace{\widetilde{H}^{(2)}}_{\text{CPU}_2} + 
\ldots +
\underbrace{\widetilde{H}^{(p)}}_{\text{CPU}_{\rm p}}}_{\texttt{MPI\_Ireduce}: \, \text{CPU}_1}
] \,, \\
\widetilde{M} = \widetilde{X}^{T} \widetilde{X} := 
\begin{bmatrix} \myunderbrace{\widetilde{X}^{{(1)}^T}}{\text{GPU}_{1}} & \myunderbrace{\widetilde{X}^{{(2)}^T}}{\text{GPU}_2}  & \ldots & \myunderbrace{\widetilde{X}^{{(p)}^T}}{\text{GPU}_{\rm p}}  \end{bmatrix}  
\begin{bmatrix}
\widetilde{X}^{(1)} \\
\widetilde{X}^{(2)} \\
\vdots \\
\widetilde{X}^{(p)}
\end{bmatrix} \hspace{-2mm}
\begin{array}{l}
\} \text{\scriptsize GPU}_{1} \\
\} \text{\scriptsize GPU}_{2} \\
\vspace{3mm}  \\
\} \text{\scriptsize GPU}_{\rm p} \\
\end{array} 
=
[
\underbrace{\widetilde{X}^{{(1)}^T} \widetilde{X}^{(1)}}_{\text{GPU}_1} + 
\underbrace{\widetilde{X}^{{(2)}^T} \widetilde{X}^{(2)}}_{\text{GPU}_2} + 
\ldots +
\underbrace{\widetilde{X}^{{(p)}^T} \widetilde{X}^{(p)}}_{\text{GPU}_{\rm p}}
]  \nonumber \\
:= [
\underbrace{\underbrace{\widetilde{M}^{(1)}}_{\text{CPU}_1} + 
\underbrace{\widetilde{M}^{(2)}}_{\text{CPU}_2} + 
\ldots +
\underbrace{\widetilde{M}^{(p)}}_{\text{CPU}_{\rm p}}}_{\texttt{MPI\_Ireduce}: \, \text{CPU}_1}
] \,, 
\end{align}
where the matrix-matrix multiplication $\widetilde{H}^{(k)} = \widetilde{X}^{{(k)}^T} \widetilde{Y}^{(k)}$ and $\widetilde{M}^{(k)} = \widetilde{X}^{{(k)}^T} \widetilde{X}^{(k)}$ is performed by  GPU$_{\rm k}$, $k \in \{1, 2, \ldots, p \}$ using the \texttt{cublasZgemm/cublasDgemm} routine, then the resultant matrix is transferred to CPU$_{\rm k}$. The additions are performed on the CPUs using the \texttt{MPI\_Ireduce} routine, reducing to CPU$_1$.

%%%%%%%%%%%%%%%%%%%%%%%%%%%%%%

\subsection{Subspace diagonalization}

The matrices $\widetilde{H}$ and $\widetilde{M}$ are first transferred from  CPU$_1$ to GPU$_1$. Next, the subspace generalized eigenproblem
\begin{align}
 \widetilde{H} \widetilde{Z} =  \widetilde{M} \widetilde{Z}  \widetilde{D}   \,,
 \end{align}
where $\widetilde{Z}$ is the matrix of eigenvectors and $\widetilde{D}$  is a diagonal matrix of the eigenvalues, is solved on GPU$_1$ using the \texttt{cusolverDnZhegvd/cusolverDnDsygvd} routine. Thereafter, the matrices $\widetilde{Z}$ and $\widetilde{D}$ are transferred from GPU$_1$ to CPU$_1$, and then from CPU$_1$ to all CPU threads using the \texttt{MPI\_Bcast} routine.  Note that \texttt{cusolverDnZhegvd/cusolverDnDsygvd} are single-GPU routines and their multi-GPU versions are currently not available, which limits the size of the eigenproblem that can be solved to $\sim$15000 orbitals, due to memory constraints. However, this does not pose a problem in the majority of practical applications, which typically target systems of 1,000 atoms or less, in AIMD calculations in particular.

%%%%%%%%%%%%%%%%%%%%%%%%%%%%%%

\subsection{Rotation}
The matrices $\widetilde{X}$ and $\widetilde{Z}$ are first transferred from the CPUs to the GPUs,  the entire $\widetilde{Z}$ is transferred from each CPU$_k$ to GPU$_k$. Next, the approximate eigenvectors of the Hamiltonian $H$ are calculated as follows:
\begin{align}
X = \widetilde{X} \widetilde{Z} := 
\begin{array}{r}
 \text{\scriptsize GPU}_{1} \{ \\
\text{\scriptsize GPU}_{2}  \{  \\
\vspace{3mm}  \\
\text{\scriptsize GPU}_{\rm p}  \{ 
\end{array} \hspace{-2mm}
\begin{bmatrix}
\widetilde{X}^{(1)} \\
\widetilde{X}^{(2)} \\
\vdots \\
\widetilde{X}^{(p)}
\end{bmatrix} 
\widetilde{Z}
= 
\begin{bmatrix}
\widetilde{X}^{(1)} \widetilde{Z} \\
\widetilde{X}^{(2)} \widetilde{Z} \\
\vdots \\
\widetilde{X}^{(p)} \widetilde{Z}
\end{bmatrix} \hspace{-2mm}
\begin{array}{l}
\} \text{\scriptsize GPU}_{1}  \\
\} \text{\scriptsize GPU}_{2}   \\
\vspace{3mm}  \\
\} \text{\scriptsize GPU}_{\rm p}  
\end{array} 
:=
\begin{bmatrix}
{X}^{(1)}  \\
{X}^{(2)}  \\
\vdots \\
{X}^{(p)} 
\end{bmatrix} \hspace{-2mm}
\begin{array}{l}
\} \text{\scriptsize GPU}_{1}  \\
\} \text{\scriptsize GPU}_{2}   \\
\vspace{3mm}  \\
\} \text{\scriptsize GPU}_{\rm p}  
\end{array} 
\end{align}
where the matrix-matrix multiplication ${X}^{(k)}  = \widetilde{X}^{(k)} \widetilde{Z}$ is performed by GPU$_{\rm k}$, $k \in \{1, 2, \ldots, p \}$, using the \texttt{cublasZgemm/cublasDgemm} routine. Thereafter, the matrix $X$ is transferred from the GPUs to the CPUs. Finally, the matrix $X$ is redistributed as follows:
\begin{align}
X: 
\begin{array}{l}
\text{\scriptsize CPU}_{1} \{ \\
\text{\scriptsize CPU}_{2} \{ \\
\vspace{3mm}  \\
\text{\scriptsize CPU}_{\rm p} \{ \\
\end{array} \hspace{-2mm}
\begin{bmatrix}
{X}^{(1)} \\
{X}^{(2)} \\
\vdots \\
{X}^{(p)}
\end{bmatrix} 
\longrightarrow
\begin{bmatrix} 
\myunderbrace{{X}_{(1)}}{\text{CPU}_{1}} & \myunderbrace{{X}_{(2)}}{\text{CPU}_2}  & \ldots & \myunderbrace{{X}_{(p)}}{\text{CPU}_{\rm p}}  
\end{bmatrix}  \,.
\end{align}
The matrix of approximate eigenvectors $X$ so generated is used as initial guess for the subsequent SCF iteration.

%%%%%%%%%%%%%%%%%%%%%%%%%%%%%%%%%%%%%%%%%%%%%%%%%%%%%%%%%%%%%%%%%%%%%%%%%%%%%%%%%%%%%%%%%%%%%%%%%%%%

\section{Results and discussion}\label{Sec:Results}
We now study the performance of the GPU-accelerated SPARC implementation through representative examples, namely bulk molybdenum (Mo), and 12-layer (100) slab of titanium dioxide (TiO$_2$) \cite{sahoo2022ab}. Specifically, we consider 250, 686, and 1024-atom unit cells of Mo, with LDA \cite{Kohn1965, perdew1981} exchange-correlation functional  and $\Gamma$-point Brillouin zone integration; and 144, 324, and 576-atom unit cells of TiO$_2$ with PBE \cite{perdew1996generalized} exchange-correlation functional and  $4\times4$, $3 \times 3$, and $2 \times 2$ Monkhorst-Pack \cite{monkhorst1976special} grids for Brillouin zone integration, respectively. We perform NVK ab-initio molecular dynamics (AIMD) with Gaussian thermostat \cite{Zhang1998} at temperatures of $3000$ and $300$ K and time steps of  1 and 2 fs for the Mo and TiO$_2$ systems, respectively. In particular, we perform $\sim 10$ steps of the AIMD simulation before collecting the timings, i.e., after  the computational timings per MD step have stabilized. 

In all calculations, we employ ONCV pseudopotentials \cite{hamann2013optimized} with nonlinear core correction (NLCC) from the SPMS set \cite{shojaei2023soft}, which has 14, 12, and 6 electrons in valence for Mo, Ti, and O, respectively. In addition, we employ the  restarted Periodic Pulay mixing scheme \cite{pratapa2015restarted, Banerjee2016PeriodicPulay},  real-space Kerker preconditioning  \cite{kerker1981efficient, realspaceprecond}, and the Alternating Anderson-Richardson (AAR) \cite{suryanarayana2019alternating, pratapa2016anderson} linear solver for the Poisson equation. The Poisson equation is solved entirely on the CPUs, since it takes a very small fraction of the total time. Indeed, it can be immediately ported to the GPUs using the Laplacian-vector product routine described in Section~\ref{Subsec:ChebFilt}, but is not done  to maximize code simplicity. The number of orbitals chosen for the Mo systems: Mo$_{250}$, Mo$_{686}$, and Mo$_{1024}$ are $N_s = 2105$, $5767$, and $8606$, respectively; and for the TiO$_2$ systems:  (TiO$_2$)$_{48}$, (TiO$_2$)$_{108}$, and (TiO$_2$)$_{192}$ the numbers are $N_s = 696$, $1560$, and $2769$, respectively, as automatically determined by SPARC. The grid spacing used for the Mo and TiO$_2$ systems is 0.372 and 0.3 bohr, respectively, which translates to $N_d = 80\times 80 \times 80$, $112\times 112 \times 112$, and $144\times 144 \times 144$ finite-difference nodes for the Mo$_{250}$, Mo$_{686}$, and Mo$_{1024}$ systems, respectively; and $N_d = 58 \times 37 \times 225$, $88\times 56 \times 225$, and $117 \times 75 \times 225$ for the (TiO$_2$)$_{48}$, (TiO$_2$)$_{108}$, and (TiO$_2$)$_{192}$ systems, respectively.  Note that all numerical parameters, including grid spacing and SCF tolerances are chosen to provide  a chemical accuracy of $10^{-3}$ Ha/atom in the energy. All simulations are carried out on the Lassen supercomputer at the Lawrence Livermore National Laboratory (LLNL) \cite{LLNLwebMachines}, wherein each computational node has 4 \texttt{NVIDIA Volta V100} GPUs with 16 GB of memory each and 40 \texttt{IBM POWER9} CPU cores with a total of 256 GB of memory. We use all 40 CPU cores per computational node in CPU-only runs, with one CPU thread (MPI rank) per CPU core, and use 4 CPU cores and 4 GPUs per computational node in GPU-accelerated runs, with one CPU thread (MPI rank) per CPU core --- the configuration that was found to be most efficient.

\begin{figure}[h!]
\centering
\subfloat[Bulk molybdenum]{\includegraphics[keepaspectratio=true,width=0.49\textwidth]{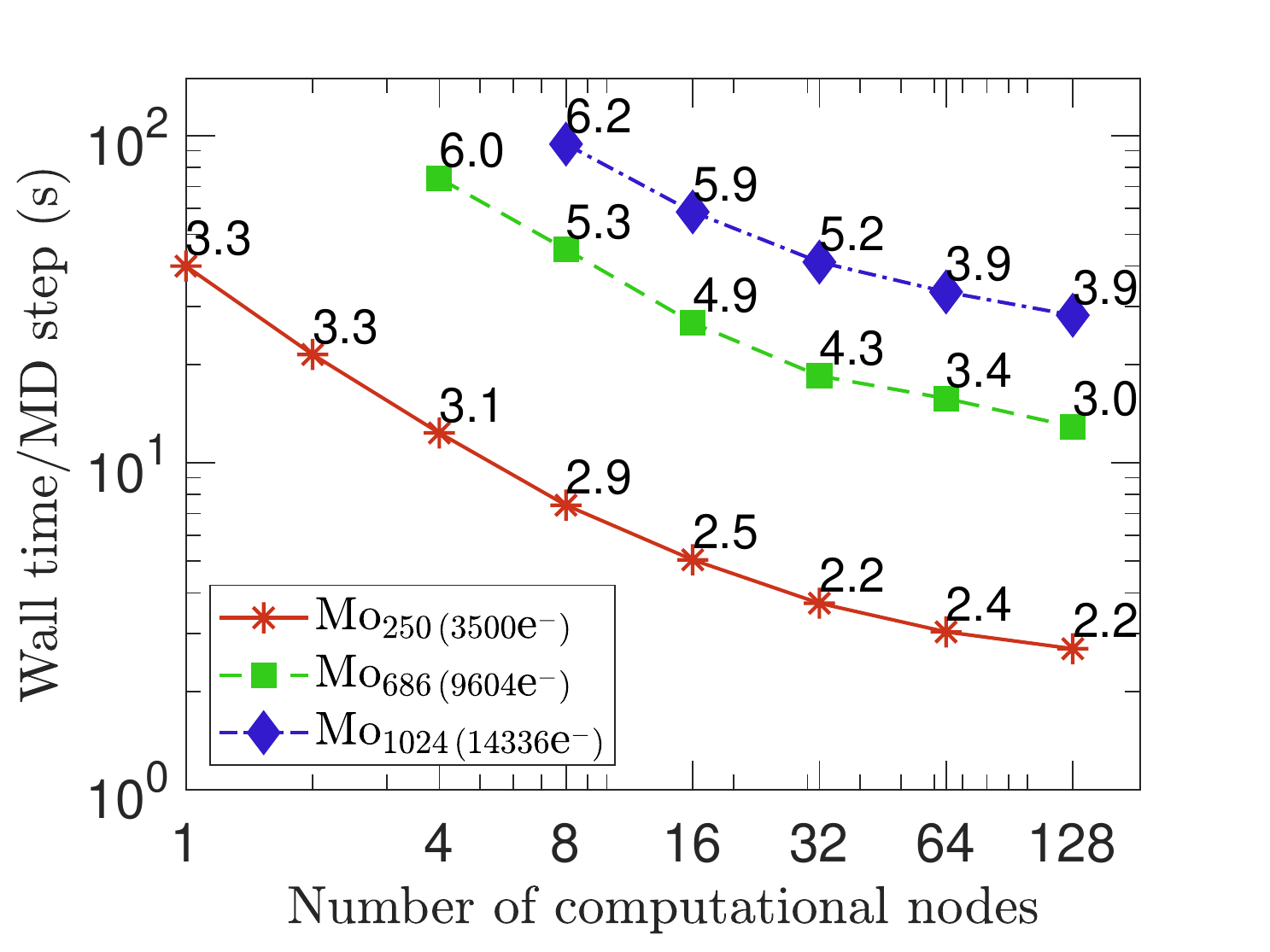} \label{Fig:Mo_scaling} }
\subfloat[(001) Titanium dioxide slab]{\includegraphics[keepaspectratio=true,width=0.49\textwidth]{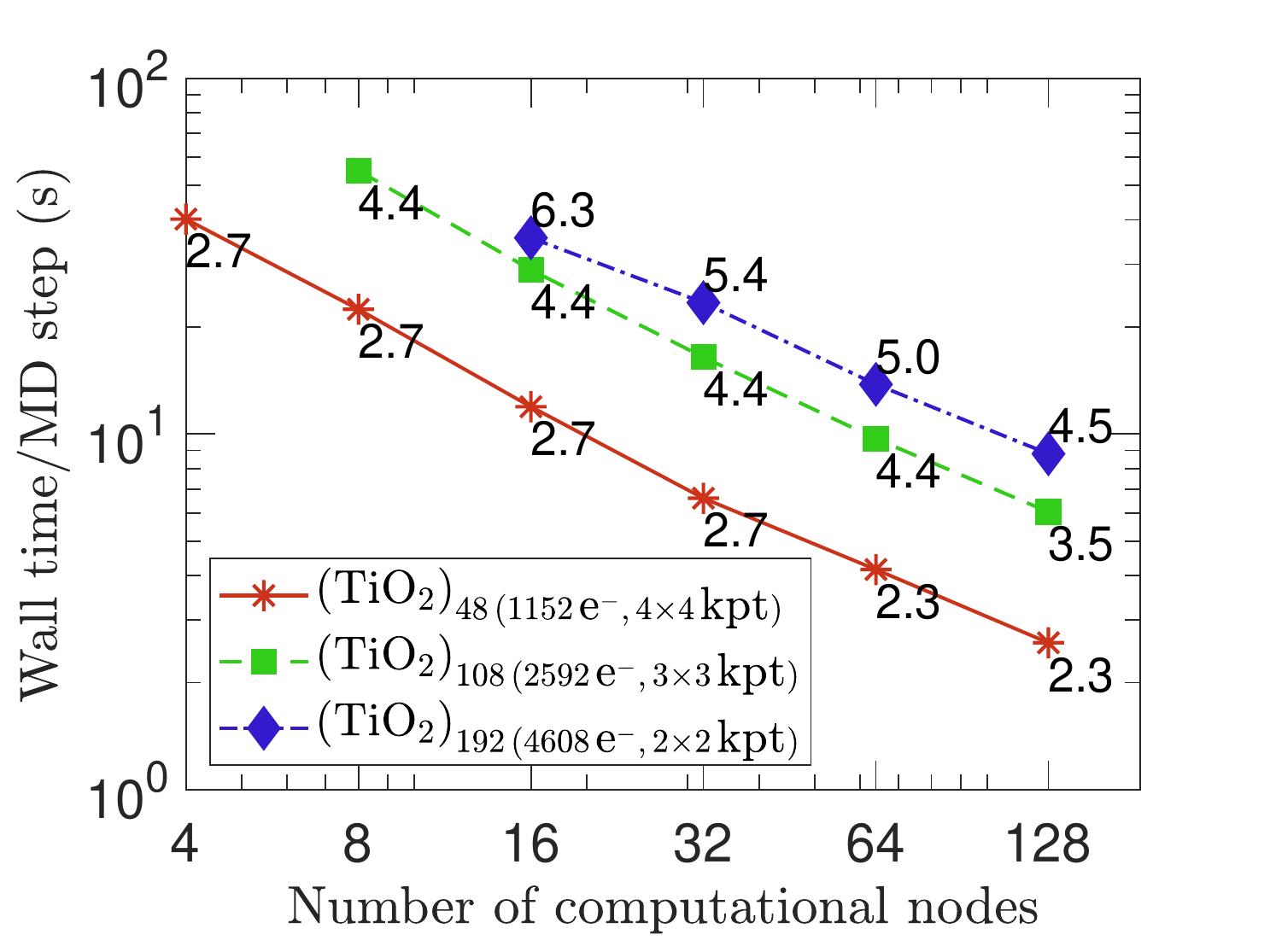} \label{Fig:TiO2_scaling} }
\caption{\label{Fig:Scaling} Strong scaling of MD step time in GPU-accelerated SPARC on the \texttt{Lassen} supercomputer \cite{LLNLwebMachines}, where each computational node has 4 GPUs and 40 CPU cores. The timings correspond to using 4 GPUs and 4 CPU threads on each computational node. The number displayed next to each marker represents the speedup in time to solution relative to CPU-only execution, wherein  all 40  CPU cores on each computational node are utilized. The number of SCF iterations per MD step for the molybdenum and titanium dioxide systems are 3 and 2, respectively. The timings include the computation of the atomic forces.}
\end{figure}

In Fig.~\ref{Fig:Scaling} we present the strong scaling results so obtained for the chosen Mo and TiO$_2$ systems. In particular, we report the variation in the total wall time per MD step --- which includes  3 and 2 SCF iterations for the Mo and TiO$_2$ systems, respectively, as well as the calculation of the Hellmann-Feynman atomic forces ---  with the number of computational nodes. It is clear that the GPU implementation demonstrates good parallel scaling, with a continuous decrease in the time to solution as the number of nodes is increased. The parallel scaling for the TiO$_2$ systems is especially good by virtue of parallelization over wavevectors in the Brillouin zone in addition to parallelization over orbitals. In particular, the GPU-accelerated execution provides significant speedup compared  to CPU-only execution, with a maximum speedup of $3.3$x, $6.0$x, and $6.2$x for  Mo$_{250}$, Mo$_{686}$, and Mo$_{1024}$, respectively; and $2.7$x, $4.4$x, and $6.3$x for (TiO$_2$)$_{48}$, (TiO$_2$)$_{108}$, and (TiO$_2$)$_{192}$, respectively. Furthermore, the minimum MD step times are well within half a minute: $2.7$, $12.9$, and $28.3$ s for Mo$_{250}$, Mo$_{686}$, and Mo$_{1024}$; and $2.6$, $6.0$, and $8.8$ s for (TiO$_2$)$_{48}$, (TiO$_2$)$_{108}$, and (TiO$_2$)$_{192}$, respectively, demonstrating the attractiveness of GPU-accelerated SPARC for performing AIMD.  The figure also indicates that the rate of decrease in the time to solution reduces as the number of computational nodes is increased, with the speedup having an inverse correlation with the number of computational nodes and a direct correlation with the problem size. This is due to the fact that within memory constraints, the GPU is able to simultaneously process much larger amounts of data in comparison to a CPU, therefore reduction in computational workload on a GPU is not in direct correspondence with the associated reduction in time. We have verified this behavior by performing the Mo$_{1024}$ simulation with a grid spacing of 0.22 bohr, e.g., corresponding to a harder pseudopotential or higher accuracy, and found a speedup and timing of $5.6$x and $86$ s on $64$ computational nodes, respectively.  The corresponding numbers for 0.372 bohr mesh (Fig.~\ref{Fig:Scaling}) are $3.9$x and $33.3$ s, respectively. Indeed, though the number of finite-difference nodes increased by a factor of $4.8$x, the wall time increased by only a factor of $2.6$x, even with the Chebyshev polynomial degree increasing from $23$ to $33$. It is worth noting that since the largest speedups occur on the smallest computational resources, the reduction in wall time is  especially useful in real-world production runs where resources are generally limited. 
%It is also worth noting that the speedups reported here correspond to the number of CPUs in CPU-only calculations being a factor of 10 larger than the number of CPUs/GPUs in corresponding GPU-accelerated calculations, as would be used in practice. Speedups would be an order of magnitude larger using the same number of CPUs in CPU-only calculations as CPUs/GPUs in corresponding GPU-accelerated calculations.
%Indeed, the speedups will be an order of magnitude larger when comparing the same number of CPUs and GPUs, e.g., $62$x for the Mo$_{1024}$ system on 32 CPUs/GPUs. 
 
\begin{figure}[h!]
\centering
\subfloat[Bulk molybdenum]{\includegraphics[keepaspectratio=true,width=0.49\textwidth]{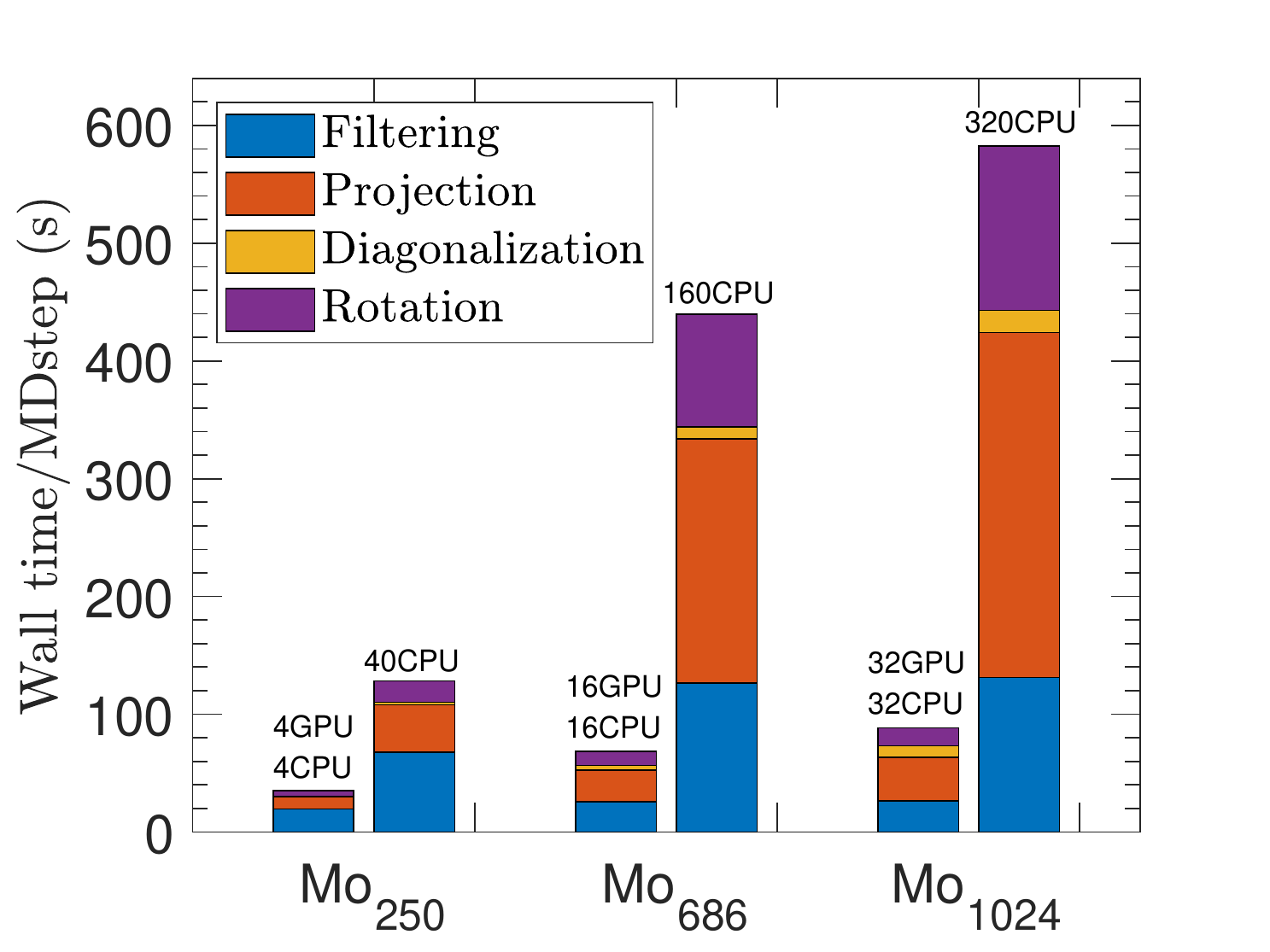} \label{Fig:Mo_timesplit_min} }
\subfloat[(001) Titanium dioxide slab]{\includegraphics[keepaspectratio=true,width=0.49\textwidth]{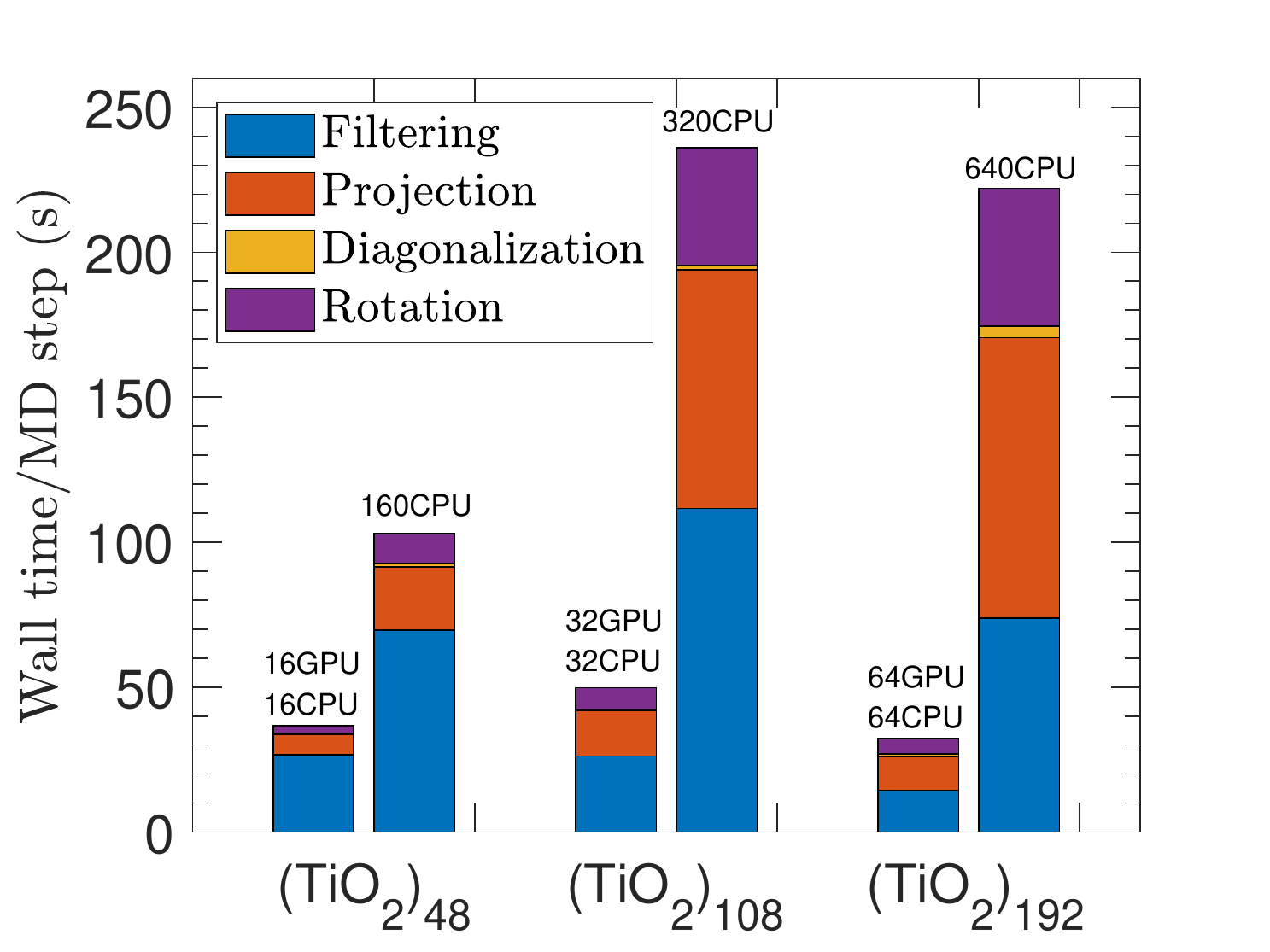} \label{Fig:TiO2_timesplit_min} }
\caption{\label{Fig:Timesplit_min} Breakdown of the timings for GPU-accelerated and CPU-only SPARC execution on the minimum number of computational nodes used in the strong scaling study (Fig.~\ref{Fig:Scaling}). 
The speedups in (filtering, projection, diagonalization, rotation) for Mo$_{250}$, Mo$_{686}$, and Mo$_{1024}$ are (3.5x, 4.0x, 3.6x, 3.8x), (4.9x, 7.7x, 2.81x, 7.9x), and (5.0x, 7.9x, 2.0x, 9.2x), respectively. The corresponding numbers for (TiO$_2$)$_{48}$, (TiO$_2$)$_{108}$, and  (TiO$_2$)$_{192}$ are (2.6x, 3.2x, 11.9x, 3.4x), (4.3x, 5.2x, 5.0x, 5.4x), and (5.1x, 8.3x, 4.1x, 8.8x), respectively.}
\end{figure}

\begin{figure}[h!]
\centering
\subfloat[Bulk molybdenum]{\includegraphics[keepaspectratio=true,width=0.49\textwidth]{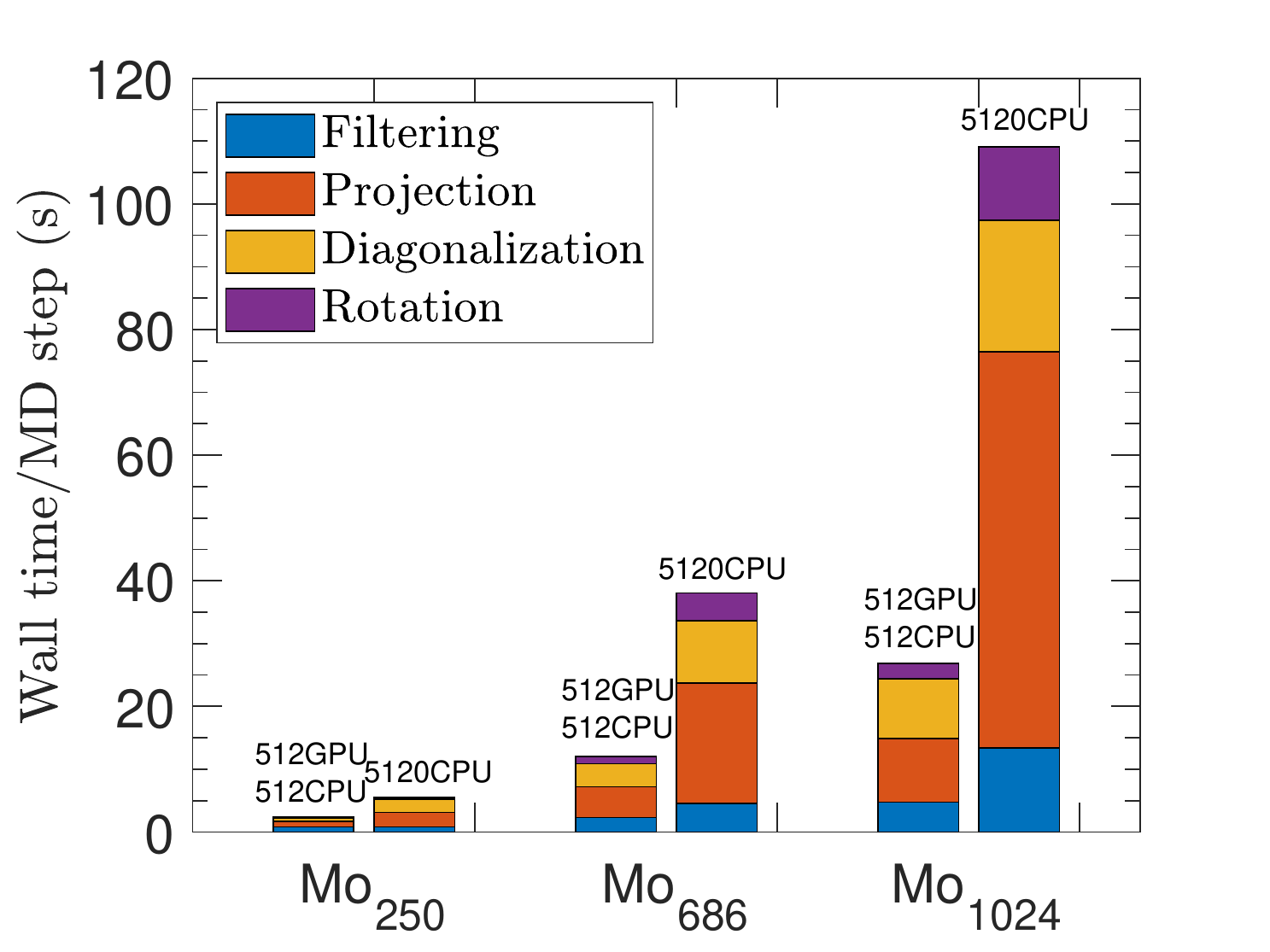} \label{Fig:Mo_timesplit_max} }
\subfloat[(001) Titanium dioxide slab]{\includegraphics[keepaspectratio=true,width=0.49\textwidth]{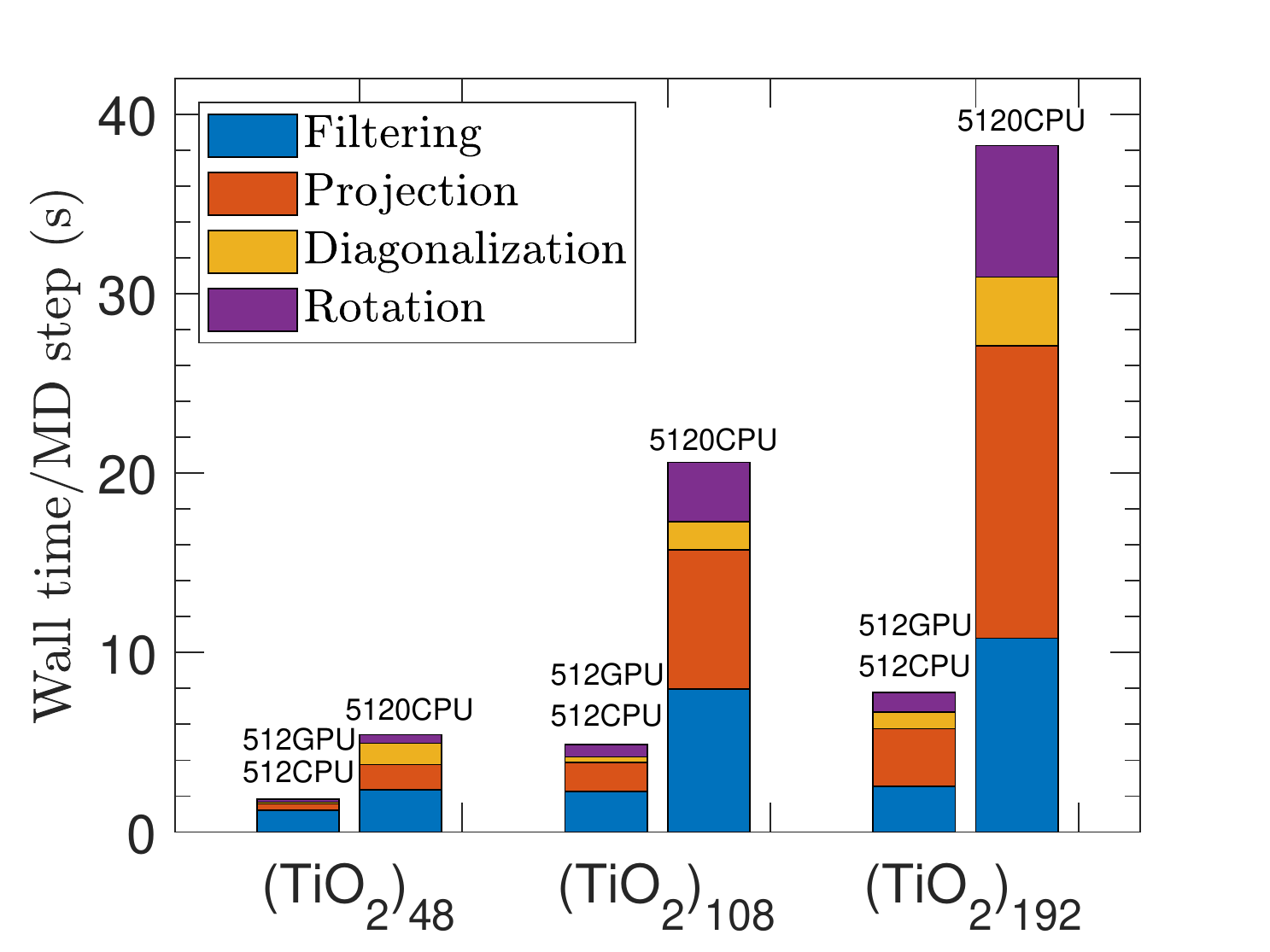} \label{Fig:TiO2_timesplit_max} }
\caption{\label{Fig:Timesplit_max} Breakdown of the timings for GPU-accelerated and CPU-only SPARC execution on the maximum number of computational nodes used in the strong scaling study (Fig.~\ref{Fig:Scaling}). The speedups in (filtering, projection, diagonalization, rotation) for Mo$_{250}$, Mo$_{686}$, and Mo$_{1024}$ are (1.0x, 2.8x, 3.6x, 1.3x), (2.0x, 3.9x, 2.81x, 3.7x), and (2.8x, 6.2x, 2.2x, 4.8x), respectively. The corresponding numbers for (TiO$_2$)$_{48}$, (TiO$_2$)$_{108}$, and  (TiO$_2$)$_{192}$ are (1.9x, 3.9x, 11.9x, 2.9x), (3.5x, 4.8x, 5.0x, 4.9x), and (4.2x, 5.1x, 4.1x, 6.7x), respectively. }
\end{figure}

To get further insight into the performance of the GPU-accelerated SPARC code, we determine the timings for each of the main CheFSI steps: Chebyshev filtering, projection, subspace diagonalization, and rotation, the details of which are available  in Section~\ref{Sec:GPUaccel}. Note that with GPU implementation of the nonlocal forces, the time taken in the calculation of the atomic forces is less than $1\%$ of the total time in GPU-accelerated execution, similar to CPU-only execution, which motivates its exclusion from the analysis here. In Figs.~\ref{Fig:Timesplit_min} and \ref{Fig:Timesplit_max}, we present the breakdown of the timings for GPU-accelerated and CPU-only executions on the minimum and maximum number of computational nodes used in the strong scaling study for each system (Fig.~\ref{Fig:Scaling}). It is clear that other than subspace diagonalization, the speedups for each of the steps on the smallest number of nodes is significantly larger than on the largest number of nodes, for the reasons discussed above with regards to the processing capability of the GPU. There is no noticeable change in the timing of the subspace diagonalization since it is restricted to a single GPU in GPU-accelerated execution, while the number of CPU threads on which it is performed is large enough in all cases that the timing remains relatively unchanged in the strong scaling study.  As is to be expected, for both GPU-accelerated and CPU-only executions,  the $\mathcal{O}(N^3)$ steps, i.e., projection, subspace diagonalization, and rotation, become more dominant as the system size increases. The strong scaling efficiency of the different CheFSI steps in GPU-accelerated execution are in the order: Chebyshev filtering $>$ rotation $>$ projection $>$ subspace diagonalization. The efficiency of the Chebyshev filtering is the highest, given that it employs orbital parallelization, whereby all the computations happen independently on the GPUs, without the need for communication between the GPUs or CPUs during the whole step. The relatively large amount of global communications that are required for forming the subspace Hamiltonian and overlap matrices during the projection step make its scaling worse than the rotation step, which would otherwise be similar. The subspace diagonalization timings remain unchanged, by virtue of being run on the same number of processors for the whole strong scaling study, as discussed above. Note that the ordering of the strong scaling efficiency of the different steps in CPU-only execution mirrors that in GPU-accelerated execution, for reasons similar to those discussed above. 

%%%%%%%%%%%%%%%%%%%%%%%%%%%%%%%%%%%%%%%%%%%%%%%%%%%%%%%%%%%%%%%%%%%%%%%%%%%%%%%%%%%%%%%%%%%%%%%%%%%%

\section{Concluding remarks} \label{Sec:Conclusions}
We have presented a GPU-accelerated implementation of the real-space SPARC electronic structure code for performing Kohn-Sham DFT calculations with LDA/GGA exchange-correlation functionals. In particular, we have developed a modular math kernel based implementation for \texttt{NVIDIA} architectures in which the computationally intensive operations are carried out on the GPUs, while the remainder of the workload is retained on the CPUs. Through representative bulk and slab examples, we have shown that GPUs enable speedups of up to 6x  relative to CPU-only execution, bringing time to solution down to less than 30 seconds for a metallic system with over 14,000 electrons, and enabling significant reductions in computational resources required for a given wall time.

The modular yet general nature of the developed implementation allows for its relatively simple extension to other GPU architectures, e.g. \texttt{AMD} and and \texttt{Intel}, which is currently being pursued by the authors. A GPU accelerated Parallel Computation Engine (libPCE) is also in development, which targets problem sizes that do not fit on a single GPU and reduces the number of CPU-GPU and GPU-CPU transfers. It uses a distinct orbital+domain data distribution, and uses CA3DMM \cite{ca3dmm} to provide optimal or near optimal communication for matrix-matrix products (used in the  CheFSI projection and rotation steps). 
% These differences require additional complexities and considerations that are beyond the scope of the current work. 
Other worthy subjects of research include extending the implementation to enable  GPU acceleration for advanced semilocal/hybrid exchange-correlation functionals, which  are significantly more computationally expensive than LDA/GGA; and GPU acceleration of the $\mathcal{O}(N)$ Spectral Quadrature (SQ) method \cite{suryanarayana2013spectral, pratapa2015spectral} in SPARC \cite{bhattacharya2021accurate, Suryanarayana2017SQDFT}, which will enable the study of systems of a million atoms \cite{gavini2022roadmap} and more as ever larger-scale parallel computing platforms become available.

%%%%%%%%%%%%%%%%%%%%%%%%%%%%%%%%%%%%%%%%%%%%%%%%%%%%%%%%%%%%%%%%%%%%%%%%%%%%%%%%%%%%%%%%%%%%%%%%%%%%
\section*{Acknowledgements}
J.E.P, A.S., and P.S. gratefully acknowledge support from U.S. Department of Energy (DOE), National Nuclear Security Administration (NNSA): Advanced Simulation and Computing (ASC) Program at LLNL, and computational resources provided under the Multiprogrammatic and Institutional Computing programs at LLNL. P.S., L.E., and E.C. gratefully acknowledge support from the U.S. Department of Energy, Office of Science under grant DE-SC0019410. 
This work was performed in part under the auspices of the U.S. Department of Energy by Lawrence Livermore National Laboratory under Contract DE-AC52-07NA27344. 
The views and conclusions contained in this document are those of the authors and should not be interpreted as representing the official policies, either expressed or implied, of the Department of Energy, or the U.S. Government.
 
\section*{Data Availability}
The data that support the findings of this study are available from the corresponding author upon reasonable request.

\section*{Author Declarations} 
The authors have no conflicts to disclose.

%%%%%%%%%%%%%%%%%%%%%%%%%%%%%%%%%%%%%%%%%%%%%%%%%%%%%%%%%%%%%%%%%%%%%%%%%%%%%%%%%%%%%%%%%%%%%%%%%%%%%
\bibliography{SPARC_GPU}
\end{document}